\documentclass[graybox]{styles/svmult}

\usepackage{mathptmx}       
\usepackage{helvet}         
\usepackage{courier}        
\usepackage{type1cm}        
\usepackage{makeidx}         
\usepackage{graphicx}        
\usepackage{multicol}        
\usepackage[bottom]{footmisc}

\usepackage[utf8]{inputenc}
\usepackage[center]{caption}
\usepackage{varioref}
\usepackage[hidelinks]{hyperref}
\usepackage[capitalise]{cleveref}
\graphicspath{{./images/}}

\makeindex             

\begin{document}

\title*{Multi Agent System for Machine Learning Under Uncertainty in Cyber Physical Manufacturing System}
\titlerunning{MAS for ML under Uncertainty in CPMS}
\author{Bang Xiang Yong and Alexandra Brintrup}
\authorrunning{Bang Xiang Yong and Alexandra Brintrup}
\institute{Bang Xiang Yong \at University of Cambridge, Institute for Manufacturing \email{bxy20@cam.ac.uk}
\and Alexandra Brintrup \at University of Cambridge, Institute for Manufacturing \email{ab702@cam.ac.uk}}
%
%

\maketitle

\abstract{Recent advancements in predictive machine learning has led to its application in various use cases in manufacturing. Most research focused on maximising predictive accuracy without addressing the uncertainty associated with it. While accuracy is important, focusing primarily on it poses an overfitting danger, exposing manufacturers to risk, ultimately hindering the adoption of these techniques. In this paper, we determine the sources of uncertainty in machine learning and establish the success criteria of a machine learning system to function well under uncertainty in a cyber-physical manufacturing system (CPMS) scenario. Then, we propose a multi-agent system architecture which leverages probabilistic machine learning as a means of  achieving such criteria. We propose possible scenarios for which our proposed architecture is useful and discuss future work. Experimentally, we implement Bayesian Neural Networks for multi-tasks classification on a public dataset for the real-time condition monitoring of a hydraulic system and demonstrate the usefulness of the system by evaluating the probability of a prediction being accurate given its uncertainty. We deploy these models using our proposed agent-based framework and integrate web visualisation to demonstrate its real-time feasibility.}

\section{INTRODUCTION} \label{section-introduction}

The use of machine learning (ML) has gained increasing popularity in a variety of manufacturing applications such as predictive maintenance, fault diagnosis, scheduling optimisation and product quality inspection \cite{wang2018deep}. To deploy these data-driven methods in industrial settings, various architectures such as multi-agent systems \cite{lee2015intelligent} and service-oriented architectures \cite{zorrilla2013service} have been proposed. 

A missing ingredient from these approaches is the measurement of uncertainty. Although manufacturing systems are characterised by heterogeneous distributed systems, dynamic and uncertain environments, currently deployed approaches focus primarily on maximising the accuracy of predictions. Without reasoning on the uncertainty of data-driven systems, manufacturers are exposed to risk in events such as sensor failures or overconfident and erroneous predictions, hindering the adoption of promising ML techniques in factory environments due to mistrust \cite{kusiak2017smart}. 

For this purpose, probabilistic modelling provides a framework for representing and manipulating the uncertainty of predictive models \cite{ghahramani2015probabilistic}. Recent research has led to development of models which are not only able to handle high-dimensional data with high accuracy but at the same time, quantify the model's uncertainty. 
Deep probabilistic models such as Bayesian Convolutional Neural Network \cite{shridhar2019comprehensive, gal2015bayesian} and Deep Gaussian Process \cite{damianou2013deep} are emerging in areas such as computer vision, autonomous vehicles \cite{mcallister2017concrete} and disease detection \cite{leibig2017leveraging}. However, there is a lack of research in quantifying and acting upon uncertainty of predictive systems in Cyber-physical Manufacturing systems (CPMS). This gap calls for a need to understand the requirement and uncertainty of a data analytics system as a core component of a CPMS. 

In \cref{section-background}, we review the relevant fields and summarise the research gap for handling uncertainty of machine learning in a CPMS, followed by establishing the performance criteria of a predictive CPMS in \cref{section-proposed_architecture}. Then, we propose a multi-agent system architecture designed to accommodate the aforementioned criteria. We present our experimental setup in \cref{section-experiment} and preliminary results in \cref{section-results}. We conclude our results and outline future research directions in \cref{section-conclusion}.

\section{BACKGROUND} \label{section-background}

In this section, we introduce the reader to the process of deploying machine learning in a CPMS, resulting sources of uncertainty in a CPMS and an overview of how these have been handled in the extant literature.

\subsection{Deploying Machine Learning in a Cyber-Physical Manufacturing System}

\begin{figure}[ht]
\centering
\makebox[\textwidth]{\includegraphics[width=\linewidth]{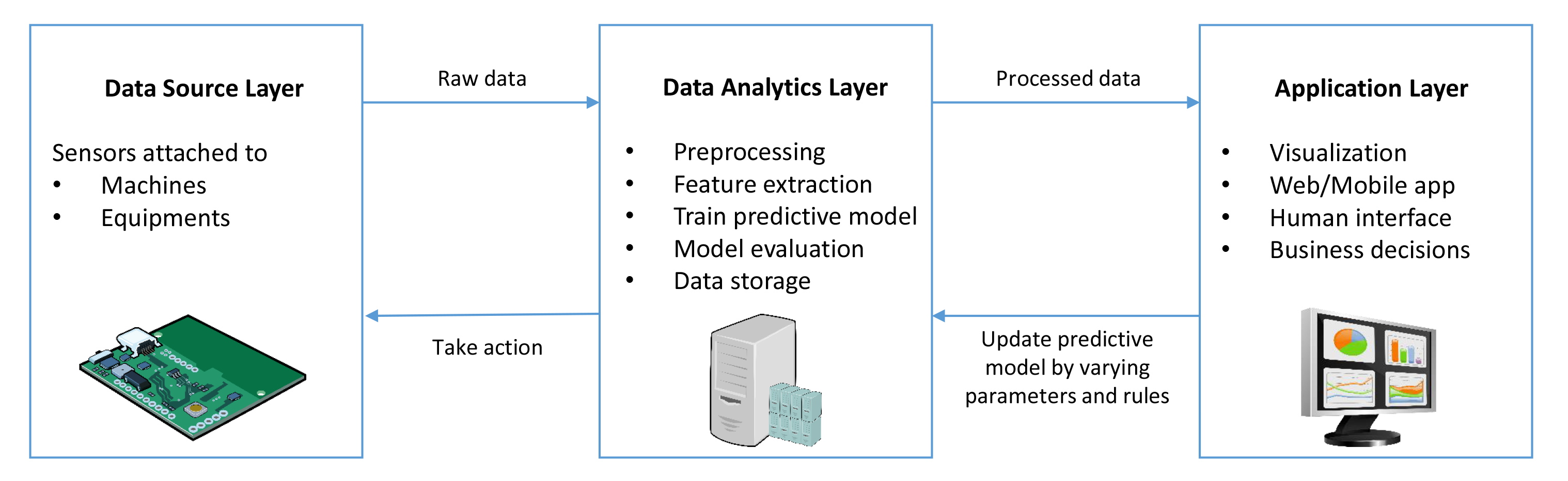}}
\caption{Layers of machine learning system in cyber-physical manufacturing system}
\label{Figure:layers}
\end{figure}

Machine learning refers to the process of extracting information and knowledge from raw data. A machine learning system is typically abstracted into three main layers: Data Source Layer, Data Analytics Layer and Application Layer. In a typical CPMS, the raw data stems from sensor measurements and usually in a time-series format and each sensor has its own sampling frequency. The data is fed to the next layer which can exist in a cloud or edge computing environment, where the actual machine learning processes are conducted. These would typically include preprocessing, extraction of features, training and evaluation of models and data storage for historical analysis. The Data Analytics Layer may provide the basis of a feedback control loop such as adding, removing, and recalibrating the sensors. In the application layer, the processed data is presented to the users in the form of an interactive visualisation. The layers which form a machine learning system in a CPMS is depicted in \cref{Figure:layers}.

In practice, as time progresses, new data is collected and as a result, the performance of the model varies. In addition, due to physical dynamics, the health and status of sensors can be affected. Consequently, in the event of sensor failures or unavailability, the input data dimension changes and existing models cannot be used. This leads to a need to monitor the predictive system and update predictive model as changes happen.

\subsection{Sources of Uncertainty in a Machine Learning Model}

\begin{figure}[ht]
\centering
\makebox[\textwidth]{\includegraphics[width=\linewidth]{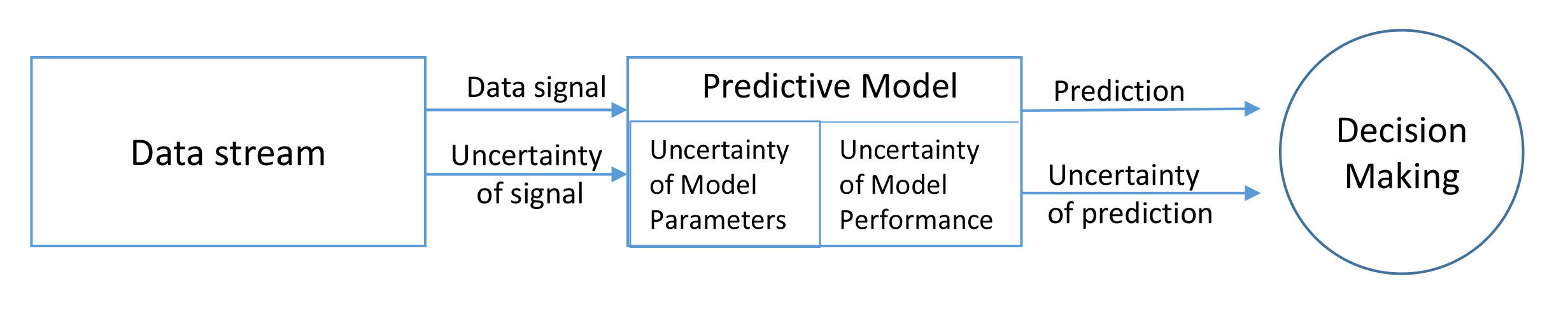}}
\caption{Uncertainty components in a CPMS predictive model}\label{Figure:Uncertainty_model}
\end{figure}

Here, we list the uncertainty pertaining components of predictive model. These are namely input signals, model parameters, prediction and performance. 
In our work, we refer uncertainty of predictive model to either one or many of these uncertainty components ( \cref{Figure:Uncertainty_model}).

\begin{enumerate}
    \item \textbf{Input signals.} Sensors always have inherent measurement uncertainty between the measured value and true value. A framework widely adopted by national metrology institutes for evaluating measurement uncertainty is the Guide to the Expression of Uncertainty in Measurement (GUM) \cite{GUM2008}. Various factors such as noise, drifts, resolution and calibration of sensors contribute to uncertainty. The associated uncertainty can either be static (given by manufacturer or calibration data) or dynamic (evaluated during runtime) \cite{eichstadt2012efficient}. 
    
    \item \textbf{Model Parameters.} The process of fitting a model to the training data involves adjusting the parameters of the model, for example, by using the weights of neural networks. Since the model parameters can take on any value to fit the data, they also have inherent uncertainty.
    
    \item \textbf{Prediction.} In addition, every prediction has an associated uncertainty with regards to the true value of what is being predicted. The prediction uncertainty is the result of two types of uncertainty: aleatoric and epistemic uncertainty \cite{gal2016uncertainty}. Aleatoric uncertainty refers to uncertainty caused by noisy data such as measurement imprecision which can be reduced by higher measurement precision. Whereas epistemic uncertainty is the uncertainty due to model parameters and structure, which can be reduced by additional data.
    
    \item \textbf{Performance.} Since the model is trained and tested on subsets of the data, the distributions of the subsets will vary. Different performance metrics are used depending on the tasks, such as mean-squared error for regression and accuracy for classification. Hence, to evaluate the model's performance on different subsets of the data, out-of-sample techniques such as k-fold validation are often used to get the mean and variance of the performance metrics. Alternatively, in-sampling techniques which exploit all available information can be used where labelling samples are expensive \cite{oneto2015performance}. 
\end{enumerate}

\subsection{Handling Uncertainty in a Cyber-Physical Manufacturing System}

Numerous studies have identified the various sources of uncertainty in CPMS. Bandyszak et al. \cite{bandyszak2018model} proposed a framework called Orthogonal Uncertainty Model (OUM) which extensively documents the uncertainty within an autonomous robot fleet. Conceptually, a taxonomy of uncertainty in CPMS has been developed in \cite{ ma2016conceptually} where uncertainty of sensor, situation, probe and knowledge can be interpreted under the uncertainty of the predictive model. 

Bayesian networks (BN) have been used widely in quantifying the uncertainty of manufacturing processes. Wolbrecht et al. \cite{wolbrecht2000monitoring} used it for monitoring multistage manufacturing process . McNaught and Chan \cite{mcnaught2011bayesian} used BN for fault diagnosis  and Nannapaneni et al. \cite{nannapaneni2016performance} used BN in welding and molding processes . However, the methods are conducted on synthetic datasets and are generally not scalable to high-dimensional inputs. In another study, Nannapaneni et al. \cite{nannapaneni2016towards} used a two-level dynamic BN for real-time control system which captures the uncertainty sources in sensors, computing resources, communication, actuation and manufacturing process. The model involves many parameters which require assumptions on the prior probability distribution and may not reflect reality. Experiments were also conducted on centralised instead of distributed systems which are common in manufacturing setting. 

Bhinge et al. \cite{bhinge2017toward} used Gaussian Process regression to develop a predictive model for energy prediction of machine tools from sensor data. Gaussian Process has also been used predicting remaining useful life of bearings \cite{hong2015}. The benefits of this method is its ability to quantify the uncertainty of predictions and the requirement of less data to train the model.

Although numerous studies have identified and quantified the uncertainties of manufacturing processes and CPMS, to this end, there is a lack of studies in quantifying and acting upon the uncertainties of a machine learning system in a CPMS. Particularly, we identify the following research gaps:
\begin{enumerate}
    \item What are the scenarios in which quantifying the uncertainty of machine learning is relevant? 
    \item What actionable insights can we gain with the quantified uncertainty of a machine learning model in a manufacturing context?
    \item How can a CPMS autonomously handle the uncertainty of the machine learning model?
\end{enumerate}

\section{Development of a Multi-Agent System for Manufacturing Machine Learning Under Uncertainty} \label{section-proposed_architecture}

Agent-based systems comprise of multiple agents which are encapsulated software processes situated in an environment and exhibiting autonomy, social ability, responsiveness and proactiveness \cite{wooldridge1995intelligent}.
In deploying predictive models, a software system is required to execute the trained model continuously on the new input data. 
For this, the usage of agent-based system for data mining has been well studied due to its inherent nature of operating autonomously in a distributed and heterogeneous environment \cite{hemamalini2014analysis}. 
In addition, its applicability in modern cloud, fog and edge computing environment has also been explored in multiple studies \cite{bakliwal2018multi,Barbosa2018ImplementationOA}. 

\subsection{Performance Criteria}
Based on extant literature, we identify important properties of a machine learning system in CPMS which need to be addressed. 
Moreover, we describe how these criteria can be linked to uncertainty of predictive models.

\begin{enumerate}
    \item \textbf{Flexibility.} This refers to the ease of modifying system components and structure during runtime \cite{kirn2006flexibility}. In our context, the addition and removal of machines, sensors, machine learning algorithms and consequent organisation of representative agents form the components of the CPMS. Such decisions can be made by analysing the impact of structural components on the uncertainty and accuracy of predictions. 
    \item \textbf{Scalability.} It is important for a system designer to estimate the impact of increasing the number of sensors, machines and complexity of models on the system's performance such as communication costs, processing time, CPU usage and memory requirement. This calls for choosing appropriate coordination mechanisms and system architecture to optimise the number of messages passed between system components and examine the influence of bottlenecks \cite{rana2000scalability}, leading to a trade-off between processing time and resolution of uncertainty of models.
    \item \textbf{Heterogeneity.} The system needs to accommodate multiple data sources and multiple predictive models for various tasks. 
    \item \textbf{Hardware Interoperability.} To encourage adoption and implementation, there is a need for interfacing with legacy and existing systems. As such, CPMS might need to be deployed in traditional centralised systems and not strictly constrained to cloud, edge and fog computing systems.
    \item \textbf{Self-Healing.} We refer to self-healing, self-adaptive and reconfiguration interchangeably as the ability to recover from and respond to dynamic events \cite{sabatucci2018four, mikic2002architectural}. By monitoring the uncertainty of predictive model, the agents can decide when to reconfigure. In the event of sensor failures, corrupted signals, or communication breakdown, the system should be capable of self-healing and tries its best to recover which ensures the availability of service. 
    \item \textbf{Safety.} Since the ground truth for each prediction may not be known immediately, the uncertainty can be used as an indicator of its accuracy. This is critical in a fail-safe scenario as it informs the user whether it is merely making a random guess or an educated guess especially in an unfamiliar environment \cite{mcallister2017concrete}. 
    \item \textbf{Interpretability.} Interpretability of machine learning reassures the user and builds trust on the system \cite{mcallister2017concrete}. There needs to be an insight into the system's prediction and uncertainty. Knowing which sensors are relevant in contributing to the models' uncertainty also improves interpretability.
    
\end{enumerate}

In light of the above requirements, Multi-agent systems appear suitable since they are widely studied for most of the criteria such as flexibility, heterogeneity, scalability and self-healing. Furthermore, the capabilities of intelligent agents for reasoning, decision making, planning and learning under uncertainty \cite{poole2017artificial} make them ideal for handling uncertainty in a CPMS. 

\subsection{PROPOSED ARCHITECTURE} 

Here we propose an agent-based architecture for incorporating machine learning under uncertainty in CPMS. In our proposed architecture, we list six agents with distinct roles: Sensor Agent, Aggregator Agent, Predictor Agent, Model Trainer Agent, Decision Maker Agent and User Interface Agent. We describe the roles and social interactions of each agent in \cref{table:Roles of Agents}. The reason for having separate agents for each task such as sensing, aggregation, prediction and decision making is to ensure the flexibility, hardware interoperability and heterogeneity of the system. 

\begin{table}
\caption{Description of system role and social interaction for agent class}
\label{table:Roles of Agents}      
%
%
\begin{tabular} {|p{2cm}| p{4cm}| p{5.3cm}|}
\hline\noalign{}
Class & System Role & Social Interaction \\ 
\noalign{}\svhline\noalign{}
Sensor Agent 
& Data source and interface to sensors in environment or machine.
& In active (passive) mode, it sends (listens) to the Aggregator Agent. 
\\ \hline

Aggregator Agent 
& Aggregates data from heterogeneous data sources.
& Interacts with the Sensor Agent for data requests and sends aggregated data to  the Predictor Agent and historical database.
\\ \hline

Predictor Agent 
& Runs the equipped predictive model on every pre-determined batch of incoming data to compute prediction and uncertainty.
& Obtains trained model from Model Trainer Agent. Handles incoming aggregated data and sends out computation result to Decision Maker Agent. 
\\ \hline

Model Trainer Agent 
& Manages, updates and deploys model. 
& Actively interacts with the Decision Maker Agent for deploying and updating model. Spawns or removes the Predictor Agent. 
\\ \hline

Decision Maker Agent 
& Obtains predictions with uncertainty and makes final decision based on uncertainty thresholds.
& Fuses computation results from Predictor Agents and actively communicates with the User Interface Agent. Sends request to Model Trainer Agent for updating or suspending Predictor Agent. 
\\ \hline

User Interface Agent 
& Acts as interface with the human user to monitor overall system and to make changes to it.
& Able to inquire status and data from agents in system based on security access level. Reads and interprets results from the Decision Maker Agent. 
\\
\noalign{}\hline\noalign{}
\end{tabular}
\end{table}
    
\begin{figure}[h]
\centering
\makebox[\textwidth]{\includegraphics[width=\linewidth]{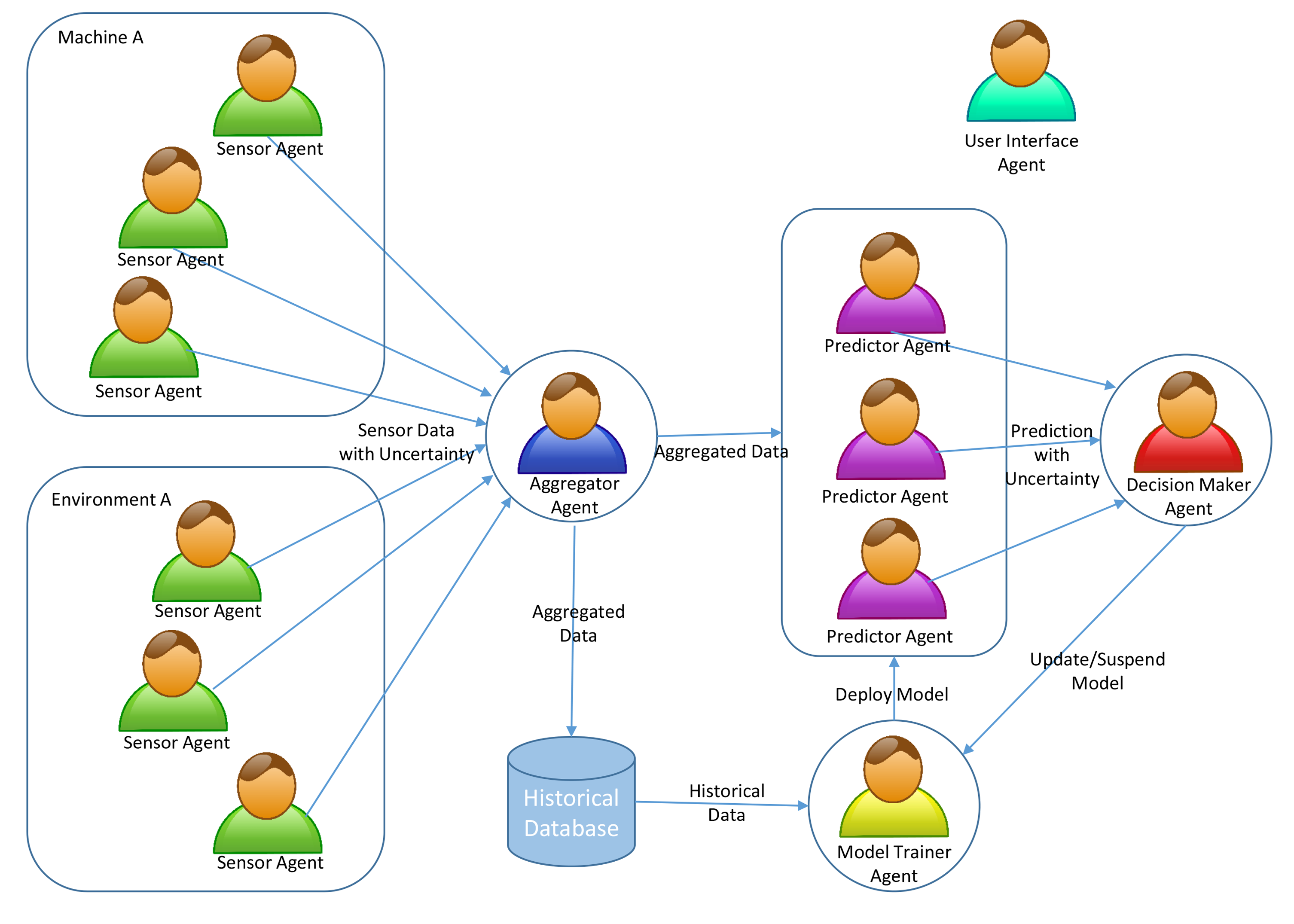}}
\caption{Proposed Architecture of Multi-Agent System}\label{Figure:MAS}
\end{figure}

Based on the social interactions among agents, we illustrate the relationship between agents in \cref{Figure:MAS}. 

In this section, we discuss the potential scenarios for our proposed architecture.

\begin{enumerate}
    \item \textbf{Handling uncertainty of predictions.} The agent-based system informs the user of the uncertainty of each prediction in real-time. This allows the user to trust the prediction when it is certain, and disregard it when uncertain, based on an acceptable threshold. Additionally, system engineers or the Decision Maker Agent can better decide on improving and maintaining the system's performance by reducing its uncertainty with provision of more data, improvement of sensor quality, or removal of sensors. 
    
    \item \textbf{Dealing with addition/removal of sensors.} 
    In an attempt to increase or decrease the number of sensors in the system, depending on the situation, new data may be collected and the existing predictive model needs to be updated. Thus, the system needs to incorporate mechanisms to automatically deal with changes in numbers of inputs to the predictive model. Such decisions may rely on evaluating the uncertainty of models. For example, if a Sensor Agent is removed and the model's uncertainty does not increase significantly, then there may be no need in retraining the model which can be expensive. 
    
    \item \textbf{Dealing with noisy or corrupted sensors.}
    The system needs to identify excessively noisy sensors which leads to high uncertainty. Alternatively, the Sensor Agent can apply suitable filter as a preprocessing step to reduce the uncertainty or zerorise the signal when necessary.
    
    \item \textbf{Decision on multiple models.} 
    There exist various methods to fuse predictions from multiple models \cite{chen2005comparative} such as majority voting. We can extend this approach by taking into account the uncertainty of each model's prediction leading to a more informed decision.  
    
    \item \textbf{Updating models.}
    Queiroz et al. \cite{queiroz2016industrial} has suggested the use of multi-agent systems to update the trained model based on system feedback. This can be extended by incorporating uncertainty of model, for example, the model can be retrained when the uncertainty of prediction is relatively high and to reduce the uncertainty of model. The agents can also explore different models which may have different uncertainty profiles and this leads to a decision between exploration and exploitation. 
\end{enumerate}

\section{EXPERIMENTAL SETUP} \label{section-experiment}

We develop a prototype multi-agent system\footnote{Code are available at https://github.com/bangxiangyong/agentMet4FoF} based on \cref{Figure:MAS} with the use case of condition monitoring of a hydraulic system. We chose Python as our programming language since many machine learning packages are available such as PyTorch \cite{pytorch2017automatic} and scikit-learn \cite{scikit-learn} which are used in our experiments. We implement the multi-agent platform using osBrain \cite{OpenSistemas2019} which is written also in Python. 

Our experiments are conducted on a real world dataset for the condition monitoring of hydraulic system \cite{schneider_tizian_2018}. The dataset consists of 2205 cycles of 17 sensors measurements with sampling frequencies of 1Hz, 10Hz and 100Hz. Each cycle lasts for 60 second. We resample all sensors into 1Hz to have a consistent sequence length for each cycle. The target variables to be classified are the conditions of the cooler, valve, internal pump, accumulator and stability. We train a separate model for each target variable. 

The analytics pipeline consists of feature extraction, normalisation and a Bayesian Neural Network (BNN). For feature extraction, we extract the mean, standard deviation, minimum, maximum, sum, median, skewness and kurtosis from both the time and frequency domains for each cycle. This results in 272 feature inputs. Then we normalise the extracted features, and feed them into the BNN. The network consists of 3 hidden layers with 272-544-272 nodes respectively. Each node's weight is represented by a probability distribution initialised as uniform distribution and trained using Bayes by Backprop \cite{shridhar2018bayesian}. The model is trained for 300 epochs, with learning rate of 0.005 and Adam optimiser \cite{kingma2014adam}. We split the data set into training and testing sets using k-fold validation, where k equals to 5. To obtain the probability distribution of each prediction, we sample the model 50 times using Monte Carlo. We obtain the modal class as the prediction class and its class percentage as the certainty measure. For evaluation of model performance, we compute the F1-score and conditional probability of a prediction being accurate given its uncertainty. Upon completing the model training and evaluation, we load the model into Predictor Agent for continuous predictions.

\begin{figure}[ht]
\centering
\makebox[\textwidth]{\includegraphics[width=\linewidth]{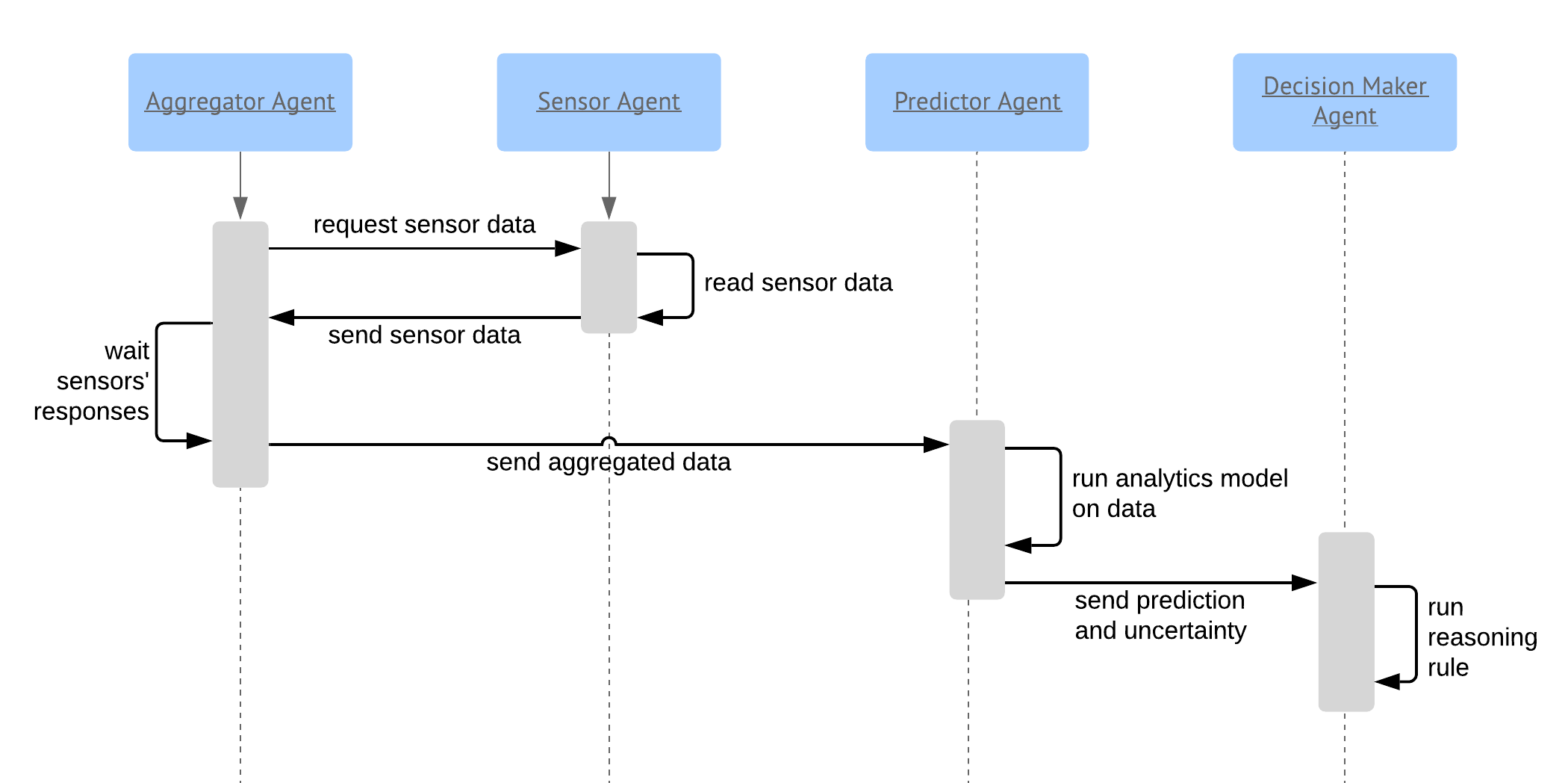}}
\caption{Sequence diagram of sensor data aggregation, prediction and decision making}\label{Figure:MAS-Sequence-Diagram}
\end{figure}

We specify the sequence diagram in \cref{Figure:MAS-Sequence-Diagram} for processing the raw data into the final prediction. The Aggregator Agent acts as a trigger by first requesting sensor data from associated Sensor Agents. It waits for all responses and aggregate them to be sent off to Predictor Agent. Then, the Predictor Agent runs the machine learning model on the received aggregated data and sends off the prediction and uncertainty to Decision Maker Agent for reasoning process. The Decision Maker Agent implements a simple filtering on the predictions based on the certainty threshold. Those above the certainty threshold, which we set as 80\%, is labelled 'certain' and below as 'uncertain'. In the background, User Interface Agent periodically fetches data from the agents to visualise the agent network relationship, sensor readings, prediction and uncertainty on a web interface.   

\section{RESULTS} \label{section-results}

For the model performance evaluation,  the results are tabulated in \cref{table:Results-performance}. Furthermore, we show that the prediction uncertainty can be used as a precursor for estimating a prediction's accuracy which is not known in run-time. For instance, for classifying the cooler's condition, given that a prediction is certain, the likelihood of being accurate is 99.70\% whereas when it is uncertain, it drops to 63.57\%. Hence, this can be used as a warning to the user whenever the prediction state is uncertain. In addition, this gives us an insight into the model's prediction and adds safety to the system in uncertain scenarios.    

\begin{table}
\caption{Model performance on condition monitoring tasks for classifying conditions of cooler, valve, internal pump, accumulator and stability}
\label{table:Results-performance}      
%
%
\begin{tabular} {|p{2.9cm}| p{2.8cm}| p{2.8cm}| p{2.9cm}|}
\hline\noalign{}
Classification Task & F1-Score & P(Accurate $\vert$ Certain) & P(Accurate $\vert$ Uncertain) \\
\noalign{}\svhline\noalign{}
Cooler Condition
& 0.99 $\pm$ 0.0002
& 99.70 $\pm$ 0.05
& 63.57 $\pm$ 5.13
\\ \hline

Valve Condition
& 0.84 $\pm$ 0.005
& 95.46 $\pm$ 0.16
& 64.69 $\pm$ 0.66
\\ \hline

Internal Pump Leakage
& 0.90 $\pm$ 0.002
& 96.56 $\pm$ 0.23
& 66.91 $\pm$ 0.45
\\ \hline

Hydraulic Accumulator
& 0.76 $\pm$ 0.01
& 95.84 $\pm$ 0.19
& 61.26 $\pm$ 1.21
\\ \hline

Stable Flag
& 0.92 $\pm$ 0.008
& 96.84 $\pm$ 0.57
& 59.92 $\pm$ 2.43
\\
\noalign{}\hline\noalign{}
\end{tabular}
\end{table}

\begin{figure}[h]
\centering
\makebox[\textwidth]{\includegraphics[width=\linewidth]{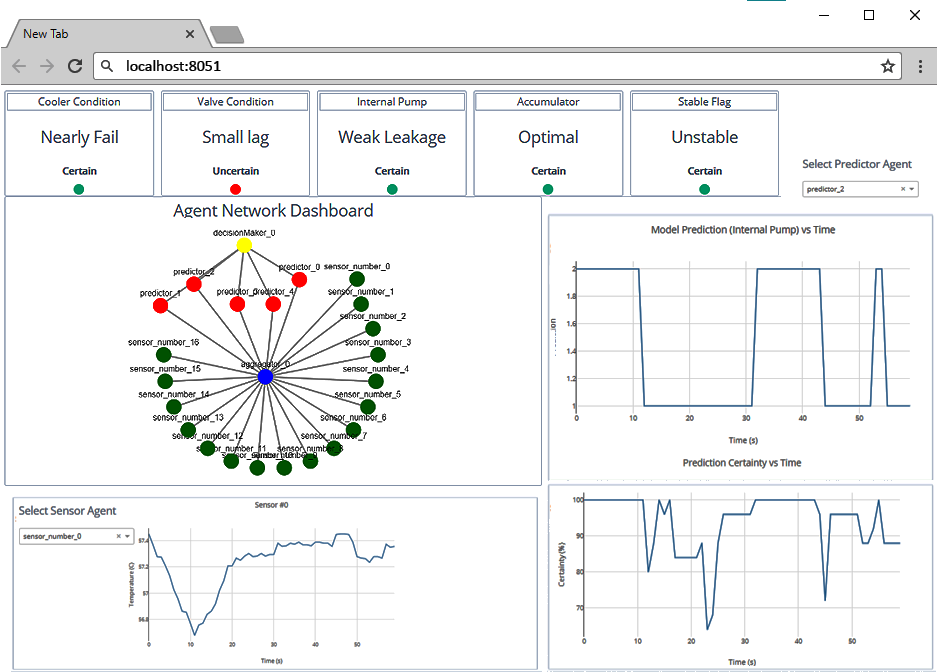}}
\caption{Screenshot of web interface for testing multi-agent system} \label{Figure:MASWeb}
\end{figure}

The interactive web interface is depicted in \cref{Figure:MASWeb}. 
The agents are displayed in a network graph along with their relationships and updated as the topology changes. As time progresses, new prediction and uncertainty are visualised in the time-series graphs, demonstrating the feasibility of deploying multiple  probabilistic machine learning models and communicating their uncertainty in our proposed architecture. 

\section{CONCLUSIONS} \label{section-conclusion}
In this paper, we explored the issue of handling machine learning uncertainty in CPMS. Firstly, we reviewed the lifecyle of a predictive model according to CRISP-DM, the deployment of machine learning in a CPMS environment. Then, we identified the sources of uncertainty in a machine learning model. These are namely the input signals, model parameters, predictions and performance. Next, we outlined the performance criteria of adopting machine learning in a CPMS under uncertainty, including flexibility, scalability, heterogeneous, hardware interoperability, self-healing, safety and interpretability. We discussed that in meeting these criteria, designing a multi-agent system based architecture appear to be ideal choice.

Hence, we proposed an agent-based architecture to deploy machine learning under uncertainty in CPMS which consists of six primary agents. We described their roles in the system and their interactions with other agents. We also specified the sequence diagram from the process of sensor data aggregation, and prediction up to decision making. 

Experimentally, a prototype of the envisioned multi-agent system was developed and tested. Using a public dataset, we implemented multiple Bayesian Neural Network models for classifying the conditions of a hydraulic system and deployed these models in the multi-agent system testbed. A web interface for visualising and managing the multi-agent system was also presented. We evaluated the performances of the proposed system and demonstrate its effectiveness of reasoning upon the uncertainty of prediction by evaluating the probability of a prediction being accurate given its uncertainty. We classified each prediction into certain and uncertain predictions based on a reasonable certainty threshold of 80\%. We find that for all tasks, the prediction accuracy is much higher given it is certain than when it is uncertain. Thus, agents can take actions to reduce the uncertainty when the current prediction is uncertain to avoid possibly erroneuous predictions. We thus conclude that our approach can be used to increase the safety and interpretability of using the machine learning system in a CPMS.

For future work, we aim to further develop, implement and analyse our proposed architecture based on the outlined performance criteria in variety of scenarios such as handling uncertainty of predictions, failure of sensors, fusing multiple models and updating models. We will further investigate the relationship between predictive uncertainty and accuracy, and how agents can make decisions to manage the uncertainty. Lastly, we aim to deploy and test in a distributed data environment such as cloud and fog computing environment.

\begin{acknowledgement}
The research presented was supported by European Metrology Programme for Innovation and Research (EMPIR) under the project Metrology for the Factory of the Future (Met4FoF), project number 17IND12. The authors thank the project partners for their valuable inputs especially Björn Ludwig and Sascha Eichstädt.
\end{acknowledgement}

\bibliographystyle{unsrt}
\bibliography{bibliography.bib}

\end{document}